\documentclass[aps,twocolumn,aps,english,showpacs,longbibliography,superscriptaddress]{revtex4-2}
\usepackage{natbib}

\usepackage{graphicx}
\usepackage{dcolumn}
\usepackage{bm}
\usepackage{color}
\usepackage{tabularx}
\usepackage{amsmath,amsfonts,amsthm,amssymb}
\usepackage{appendix}
\usepackage[english]{babel}
\usepackage{mathptmx}
\usepackage{graphicx}
\usepackage{xcolor}

\usepackage[colorlinks=true,linkcolor=blue,citecolor=blue,anchorcolor=blue,urlcolor=blue]{hyperref}
\usepackage{xr}
\makeatletter
\newcommand*{\addFileDependency}[1]{
  \typeout{(#1)}
  \@addtofilelist{#1}
  \IfFileExists{#1}{}{\typeout{No file #1.}}
}
\makeatother

\begin{document}

\title{Chiral Phonons Enable Ultrafast Magnetization Switching via Magnetoelastic Coupling}

\author{Weiwei He}
\affiliation{State Key Lab of Mechanics and Control for Aerospace Structures \& Key Lab for Intelligent Nano Materials and Devices of Ministry of Education, Nanjing University of Aeronautics and Astronautics (NUAA), Nanjing 210016, China}
\affiliation{Instituto de Ciencia de Materiales de Madrid (ICMM), CSIC, Madrid 28049, Spain}
\author{Muhammad Hamza Asim}
\author{Mara Strungaru}
\affiliation{Department of Computer Science, The University of Manchester, Manchester M13 9PL, United Kingdom}
\author{Roy Chantrell}
\affiliation{School of Physics, Engineering and Technology, University of York, York YO10 5DD, United Kingdom}
\author{Min Yi}
\email{yimin@nuaa.edu.cn}
\affiliation{State Key Lab of Mechanics and Control for Aerospace Structures \& Key Lab for Intelligent Nano Materials and Devices of Ministry of Education, Nanjing University of Aeronautics and Astronautics (NUAA), Nanjing 210016, China}
\author{Oksana Chubykalo-Fesenko}
\email{oksana@icmm.csic.es}
\affiliation{Instituto de Ciencia de Materiales de Madrid (ICMM), CSIC, Madrid 28049, Spain}

\begin{abstract} 
Phonons are desirable for current-free energy-efficient spin manipulation and harnessing their chirality to achieve ultrafast magnetization switching remains actively pursued. Here we demonstrate that terahertz-driven chiral phonons enable faster magnetization switching than linear phonons through the purely magnetoelastic coupling that transfers both energy and angular momentum. Only one phonon handedness efficiently transfers angular momentum to the spin system, which we explain by a large fictitious kinematic Barnett‑like field at THz frequencies that destabilizes spin precession for the opposite chirality. Considering different regions of the spinwave spectrum, we find that switching conditions can be realized near the $\Gamma$-point where strong energy transfer and spinwave excitation dominate, and near the $P$-point which offers minimal energy loss. Our results establish phonon chirality as a decisive and previously overlooked parameter in spin-lattice dynamics within the magnetoelastic coupling mechanism, offering a promising avenue for ultrafast low-energy spintronic devices.

\end{abstract}

\maketitle
The majority of research on ultrafast magnetization switching has focused on metallic systems~\cite{Kirilyuk2010Ultrafast,Scheid2022Light}, with ferrimagnetic GdCoFe alloys and multilayers~\cite{Ostler2012Ultrafast,Scheid2022Light,Kirilyuk2010Ultrafast}, being the prototype materials where switching is robust and has been confirmed in multiple experiments. The primary mechanism involves electronic excitation and scattering, leading to different spin-flip probabilities in Gd and Co, ultimately resulting in their distinct demagnetization rates and the exchange of angular momentum~\cite{Radu2011Transient, Ostler2012Ultrafast, Mentink2012Ultrafast}. Recently, the emerging area of ultrafast phonomagnetism has opened new possibilities for exciting magnetization dynamics through phonons rather than electrons~\cite{Stupakiewicz2021Ultrafast, Luo2023Large, Davies2024Phononic, Luo2025Terahertz}. The potential for magnetization switching via phonons is particularly exciting, as it paves the way for using insulators instead of metals. It is known that phonon-spin coupling is significant, especially near the crossing points of their corresponding spectral curves where polarons are formed. Even more importantly, the specific heat of phonons is larger than that of electrons suggesting that much less heating  occurs during magnetization switching events. This opens a door for ultrafast energy-efficient memories~\cite{Juraschek2021Magnetic}.  However, while the transfer of angular momentum between spins and phonons at the ultrafast timescale is well established~\cite{Dornes2019The, Mankovsky2022Angular, Luo2023Large}, experiments on phonon-induced magnetization switching~\cite{Stupakiewicz2021Ultrafast,Gidding2023Dynamic,Davies2024Phononic} remain scarce, and the underlying mechanisms and conditions are currently under debate.

In parallel, chiral phonons and their interaction with spin systems have garnered considerable attention~\cite{Juraschek2025Chiral}. The transfer of angular momentum between spins and chiral phonons is fundamental to the Einstein--de Haas and Barnett effects~\cite{Einstein1915Experimental,Barnet1915Magnetization}, as well as to their ultrafast counterparts~\cite{Dornes2019The, Davies2024Phononic}.  Recently, chiral phonon trajectories excited by circularly-polarised THz lasers, which plays an important role in understanding the above effects, have been experimentally measured~\cite{Minakova2026Observation}.  Recent theories predict very large transient effective fields of quantum-mechanical origin, generated by chiral phonons in magnetic systems due to charge motion with magnitude reaching up to 100\,T~\cite{Juraschek2022Giant}. These  ideas have recently been used to interpret experiments showing the polarization of paramagnetic materials~\cite{Luo2023Large} and even stimulating or impeding magnetization switching driven by circularly-polarised laser excitation~\cite{Davies2024Phononic}. However, experimental estimates of the effective magnetic fields~\cite{Luo2023Large, Davies2024Phononic} are  much smaller than the theoretically predicted values.

On the other hand, the field of magneto-acoustics has a history spanning over 70 years~\cite{Schlomann1960Generation, Gurevich1996Magnetization}, but it has recently gained new momentum by demonstrating the ability to influence magnetization dynamics through surface acoustic waves created by contact of ferromagnetic thin films with piezoelectric substrates~\cite{Li2021Advances, Vlasov2020Magnetization,Rovirola2025}.  The magneto-elastic coupling resulting in additional magnetic anisotropy is the core effect in magneto-acoustics, adequately explaining most of the observed phenomena. This mechanism has also been predicted to enable fast~\cite{Yi2015Mechanically} and ultrafast~\cite{Kovalenko2013New} switching  and has been invoked to explain the observed THz phonon- and stress-induced magnetization switching through magnetoelastic fields in terms of micromagnetics~\cite{Stupakiewicz2021Ultrafast,Yi2019Strain}.  It seems clear that this mechanism should work with linearly-polarised phonons and any role of their chirality is unclear.

Thus, determining which of the above mechanisms is responsible for the observed THz phonon-induced magnetization dynamics remains an open question. Although chiral phonons can be excited by circularly-polarised laser pulses, it is unclear whether the associated quantum-mechanical fields are sufficiently strong to induce magnetization switching. Conversely, while magnetoelastic coupling is invariably present, the extent to which phonon chirality contributes to this mechanism has yet to be uncovered.

Here, we perform spin-lattice atomistic dynamics to investigate the possibility of phonon-enabled ultrafast magnetization switching via a magnetoelastic mechanism. We explore different points in the phonon spectrum and demonstrate that magnetization can be switched by phonon excitations at different points of the spinwave spectra, e.g., near the $P$- and $\Gamma$-points with both linear and chiral phonons, albeit only phonons with one chirality are effective. However, simulations show that chiral phonons transfer angular momentum more rapidly, leading to faster switching than linear phonons. Most importantly, the switching near the $P$-point is found "cold", meaning no energy lost in spin waves, but significant energy release in terms of spin waves occurs near the $\Gamma$ point where the two spectra cross.

The total Hamiltonian of the coupled system (more details in the Supplementary Information and Refs.~\cite{Strungaru2021Spin,Strungaru2024Route}) is given by the sum of the lattice,  magnetic and coupling parts, where the lattice part is
\begin{equation}
  \mathcal{H}_\mathrm{lat} = \sum_i \frac{m_{\rm at} \mathbf{v}_i^2}{2} + \sum_{i<j} U(r_{ij}),
\end{equation}
$m_{\rm at}$ is the atomic mass, $\mathbf{v}_i$ is the atom velocity, $r_{ij}$ is the inter-atomic distance and $U(r_{ij})$ is the  interatomic potential. The spin part accounts for exchange interaction and uniaxial anisotropy term, i.e.,
\begin{equation}
    \mathcal{H}_\mathrm{mag} = -\sum_{i,j} J(r_{ij})(\mathbf{s}_i\cdot\mathbf{s}_j) - \sum_{i}K(\mathbf{s}_i\cdot \hat{\mathbf{z}})^2.
\end{equation}
The exchange term considers interaction up to the six neighbors and is distance-dependent following $J(r_{ij}) = J_0(1 - r_{ij}/r_c)^3 \Theta(r_c - r_{ij})$, where $r_c$ is the cut-off distance and $ \Theta$ is the Heaviside function.  
The coupling term~\cite{Akhiezer1968Spin,Beaujouan2012Anisotropic} is
\begin{equation}
\mathcal{H}_\mathrm{c} =  - \sum_{i,j} f(r_{ij}) \left[ (\mathbf{s}_i\cdot\hat{\mathbf{r}}_{ij})(\mathbf{s}_j\cdot\hat{\mathbf{r}}_{ij}) - \frac{1}{3}\mathbf{s}_i\cdot\mathbf{s}_j \right].
\end{equation}
where $f(r_{ij})=CJ_0(a_0/r_{ij})^4$ dictates the coupling strength where $a_0$ is the inter-atomic distance. 

The strength of the coupling term can be parameterized through the values of magneto-elastic anisotropy or magnetic damping~\cite{Dilina2016Reinventing,Strungaru2021Spin}. The pseudo-dipolar coupling term has been previously shown to correctly thermalise the spin system and allows for angular momentum transfer between the two sub-systems~\cite{Strungaru2021Spin}. Importantly, it originates in the dynamical crystal
field and is quadratic in both spin and atomic position variables. The magneto-elastic origin of the coupling term is easily seen if we assume a coherent magnetization dynamics, i.e.  $\mathbf{s}_i=\mathbf{s}_j$. In this case,  $\mathcal{H}_\mathrm{c} =-\sum_{\alpha,\beta}\epsilon_{\alpha\beta}(\mathbf{k})s_{\alpha}s_{\beta}$ with $\alpha, \beta=x,y,z$ and $\epsilon(\mathbf{k})$ being the phonon $\mathbf{k}$ vector-dependent strain tensor. For concreteness, we use Fe as an example, as it is one of the most well-studied materials with well-established parameters. The corresponding equations of motion consist of Newton's law for the atomic positions and the Landau-Lifshitz equation  for the spin dynamics. All parameters can be found in the Suppl. Info. S1. The initial magnetization direction is parallel to the $z$-anisotropy axis. 

\begin{figure}[!t]
\includegraphics[width=0.8\linewidth, trim={0 2cm 0 0}]{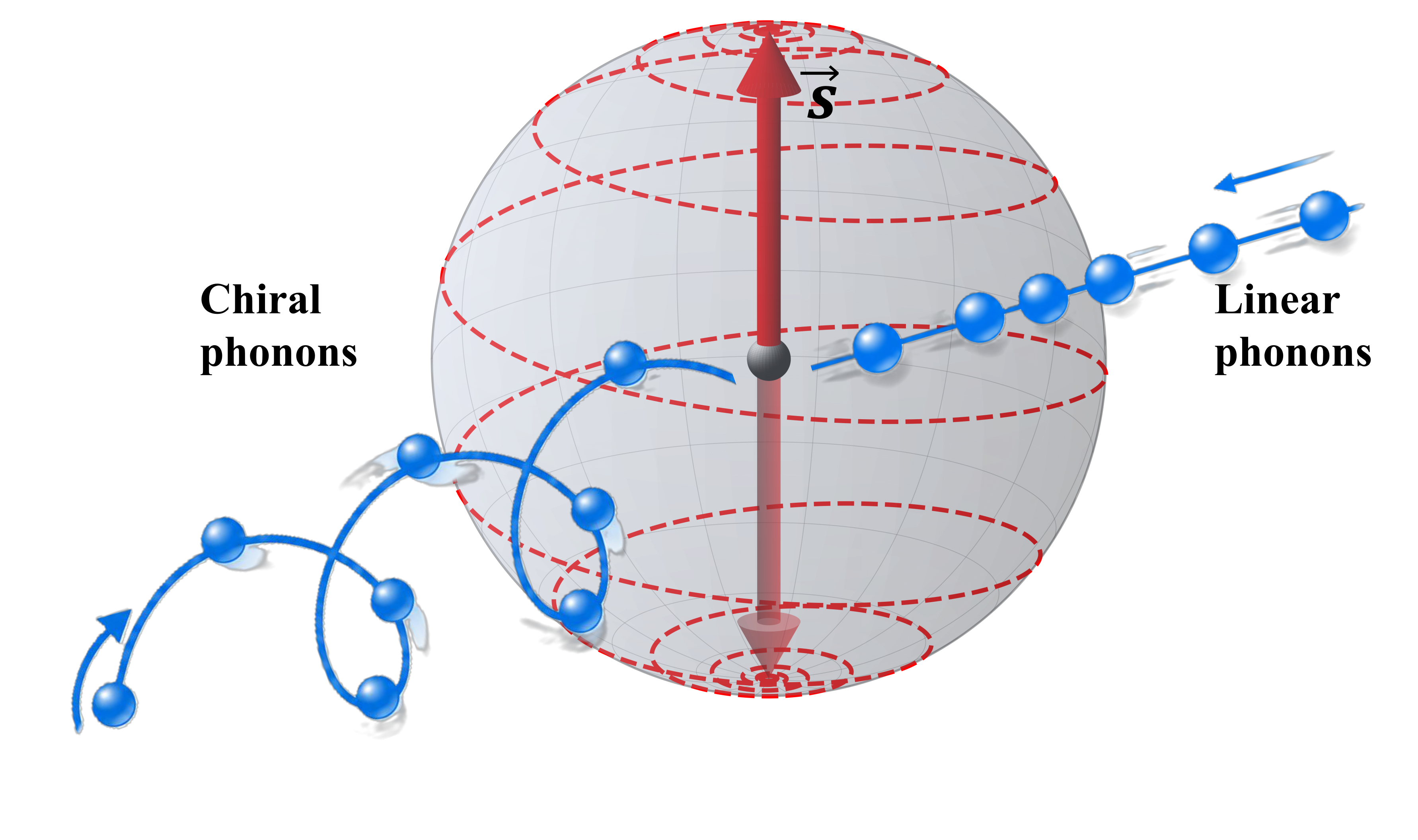}
\caption{\label{fig:illu}  Illustration of switching magnetization by left-handed chiral  or linear phonons excited at THz frequencies.}
\end{figure}

 To investigate the switching mechanism driven by different phonon modes, the external driving force $\mathbf{F}_\mathrm{THz}(t, \mathbf{r}_i)$ is applied to the atoms at specific direction for a linear excitation $\mathbf{F}_\text{linear}(t, \mathbf{r}_i) = f_0 \mathbf{e}_x \cos(2\pi \nu t - \mathbf{k}\cdot\mathbf{r}_i) \Theta(t_p - t),$
where $f_0$ is the excitation strength, $\nu$ is the excitation frequency,  $\mathbf{k}$ is the wave vector and $t_p$ is the pulse duration.
To excite chiral phonons, e.g. in the $xz$-plane, we add an orthogonal force component with a phase difference of $\pm\pi/2$ to the above linearly-polarised expression, i.e., $\pm f_0 \mathbf{e}_z \sin(2\pi \nu t - \mathbf{k}\cdot\mathbf{r}_i) \Theta(t_p - t)$,
where $\pm$ sign dictates the chirality (right-handed for positive angular momentum $L_y>0$ or left-handed for negative angular momentum $L_y<0$, respectively). This enables the direct injection of a mechanical angular momentum into the lattice. As illustrated in Fig.~\ref{fig:illu}, we can excite both linear and chiral phonons at THz frequencies and follow the evolution of the total reduced magnetization $\mathbf{m}\equiv\mathbf{M}/M_s=\frac{1}{N}\sum_i\mathbf{s}_i$
($\mu_s$ is atomic magnetic moment and $N$ is the total spin/atom number). In this article we present results with phonons excited by $k$-vector along [111] direction as an illustrative example. 
 
To explicitly trace the angular momentum transfer and energy dissipation during the switching events, we define the lattice angular momentum ($\mathbf{L}_\mathrm{ph}$ in units $\hbar/{\rm atom}$) and the spin temperature~\cite{Ma2010Temperature} (the measure of the energy in the spin system) :

\begin{equation}
\mathbf{L}_\mathrm{ph} = \frac{1}{N\hbar}\sum_{i} m_{\rm at} \left( \mathbf{r}_i \times \mathbf{v}_i \right)\, ,\;  T_\mathrm{s} = \frac{\mu_s}{2 k_B} \frac{\sum_{i} \left| \mathbf{s}_i \times \mathbf{H}_i \right|^2}{\sum_{i} \mathbf{s}_i \cdot \mathbf{H}_i},
\end{equation}
where $\mathbf{H}_i$ is the internal field acting on spin $\mathbf{s}_i$ and the index $i$ labels atoms/spins. More definition details can be found in Suppl. Info. S2.

\begin{figure}[!t]
\includegraphics[width=\linewidth]{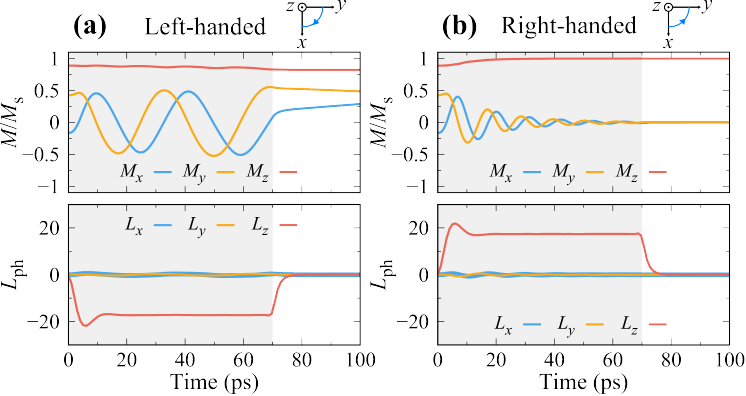}
\caption{\label{fig:Pxy} Excitation with chiral phonons excited in $xy$-plane near $P$-point. Components of the reduced magnetization $M_\alpha/M_s$ and the phonon angular momentum $L_\alpha$ are presented as a function of time, where $\alpha=x, y, z$. 
To avoid the stacking in the metastable state, a small initial random deviation of spins from $s_z=1$ direction was introduced.  (a)\,$f_0^x>0,f_0^y<0$,left-handed and (b)\,$f_0^x>0, f_0^y>0$, right-handed chiral phonon. Both phonon modes are excited with a force strength of $f_0 = 0.06$ and a frequency of 8.3\,THz. The gray region shows the interval for the THz pulse duration.}
\end{figure}

In Fig.~\ref{fig:Pxy} we illustrate an efficient excitation of precession  with chiral phonons driven in $xy$-plane. As an example, we excite phonon mode near the $P$-point of the phonon spectrum, see Suppl. Info. Fig.~S1 for the complete spectrum. It is important to note that the $P$-point is close to the edge of the Brillouin zone where the phonon branches are degenerate (longitudinal and transverse phonons have the same frequency) and thus the excitation along cartesian coordinates produces eigenmodes.  It is straightforward to see that excitation in the 
$xy$-plane is associated with phonon  angular momentum  parallel or anti-parallel to the easy-axis $z$. The resulting magnetization dynamics shows 
magnetization precession with both phonon polarisations, albeit in one case the precession  is rapidly damped, while in the other it is amplified over time. This is also seen  in the dynamics of the $z$-component of magnetization which is decreasing or increasing with time. We conclude that only the phonon with $L_z<0$, 
excites the system efficiently.    

 Phonon-induced magnetization precession can be qualitatively understood considering a simplified model of coherent magnetization dynamics when $\mathbf{s}_i=\mathbf{m}$ for all $i$ (see more details in Suppl. Info. S5). For the excitation in the $xy$-plane near the $P$-point, this leads to oscillating  elements of the stress matrix $\epsilon_{xx}$, $\epsilon_{yy}$ and  non-diagonal elements $\epsilon_{xy}=\epsilon_{yx}$. This produces rotating clock-wise or counter clock-wise fields in the $xy$-plane.
In the rotating system of coordinates around the $z$-axis, a fictitious inertia field  $H_{B}= \pm 2\pi\nu/\gamma$, directed along phonon angular momentum (analogous to the Coriolis force and  known as \emph{Barnett} field in spin dynamics~\cite{Ono2015Barnett}), 
changes the system equilibrium to a new stationary point, corresponding to a magnetization precession in the laboratory system. Most importantly, the field is huge (for $\nu$=8.3\,THz, $H_{B}\approx300$\,T!) meaning that at THz frequencies it  dominates the dynamics. The analysis of the precession stability with no damping shows that for right-handed phonons the precession is stable while it is unstable for the left-handed phonons. The interpretation in the laboratory system of coordinates is that if the field rotates with the magnetization, it stabilizes the precessional motion, whereas if it counter-rotates, it drives the magnetization away from the steady precessional trajectory.

\begin{figure}[!t]
\includegraphics[width=\linewidth]{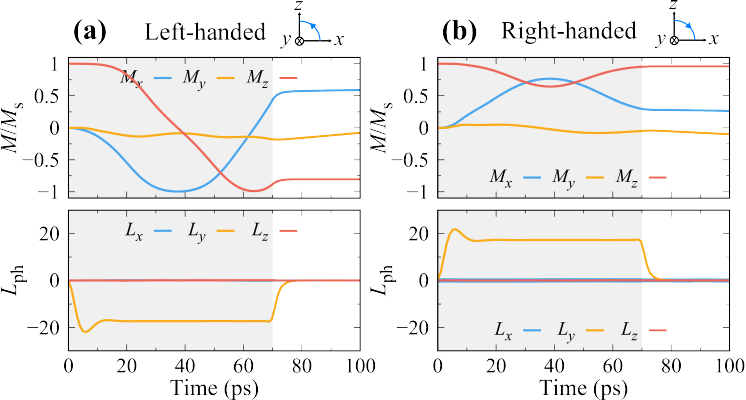}
\caption{\label{fig:Pxz} Excitation with chiral phonons excited in $xz$-plane near $P$-point. The basis vectors for the excitation plane are chosen to be $\mathbf{x}$ and $\mathbf{z}$. (a)\,$f_0^x>0, f_0^z>0$, left-handed and (b)\,$f_0^x>0,f_0^z<0$, right-handed chiral phonon. Phonon modes are excited with a force strength of $f_0 = 0.06$ and a frequency of 8.3\,THz. Reduced magnetization components and phonon angular momentum components are presented as a function of time. The gray region shows the interval for the THz pulse duration.}
\end{figure}

Conditions for the switching can be found in many situations.  In Fig.~\ref{fig:Pxz} we present an example of a successful magnetization switching driven by left-handed chiral phonons excited in the $xz$ plane with the $\mathbf{k}$-vector near the $P$-point of the phonon spectrum. In this case the phonon angular momentum has a non-zero component along the direction perpendicular to the excitation plane $L_y\neq 0$.
As before, the lattice motion  directly follows the geometry of the applied $xz$-plane excitation, resulting in an angular momentum perpendicular to this plane.  Note also an opposite chirality of the driving force and the phonon angular momentum in this case.
The right-handed phonons do not produce switching. As was previously shown~\cite{Strungaru2024Route}, the driving along the $x$-direction near the $P$-point of the spectrum (linear phonons) also leads to magnetization switching. The excitation produces magneto-elastic fields perpendicular to the initial magnetization, which in turn exert a torque and drive the dynamics.  The magnetization vector oscillates around the internal field  ending in the opposite minimum if the excitation is switched off at the right moment. With circularly-polarised phonons the switching is also precessional, as demonstrated by the development of perpendicular magnetization components. At the same time, the present results show that this energetic picture is oversimplified and insufficient to explain the results since there is a clear dependence on the phonon chirality and injected angular momentum.  We note that while the injection of  phonons with $L_z\neq 0$ mainly induce precession as shown in Fig.~\ref{fig:Pxy},  the switching shown in Fig.~\ref{fig:Pxz} is driven by the in-plane phonon angular momentum ($L_y$).

\begin{figure}[!b]
\includegraphics[width=\linewidth]{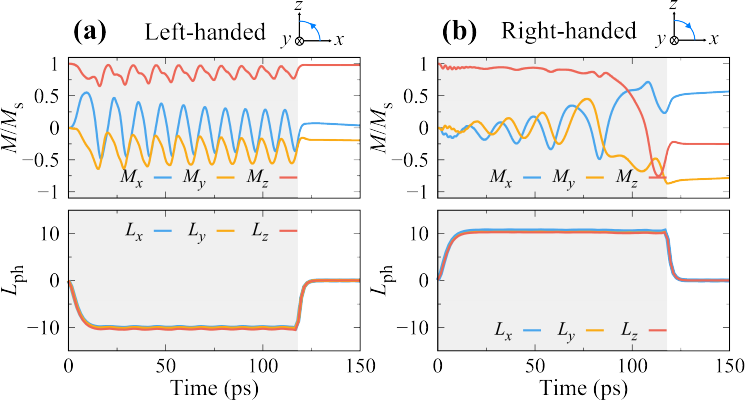}
\caption{\label{fig:Gamma-P-xz} Excitation with chiral phonons in $xz$-plane near $\Gamma$-point. (a)\,$f_0^x>0, f_0^z>0$, left-handed and (b)\,$f_0^x>0, f_0^z<0$, right-handed chiral phonon. Both phonon modes are excited with a force strength of $f_0 = 0.06$ and a frequency of 3.08\,THz. Reduced magnetization components and phonon angular momentum components are presented as a function of time. The gray region shows the interval for the THz pulse duration.}
\end{figure}

The switching mechanism with circular phonons is qualitatively similar to that of the precession excitation in the model of coherent magnetization dynamics. The Barnett field is now directed along $\pm y$ direction. This largely displaces the stationary positions in the rotating frame, producing large-amplitude magnetization precession, involving the $M_z$ component in the laboratory system. The Barnett field again stabilizes the precession for one phonon chirality and kicks it out for the opposite chirality. Together with the finite excitation duration the latter case finally determines the switching event.

Particularly interesting for the experimental observation is the phonon excitation near $\Gamma$-point of the spectrum, i.e. with long wavelength and especially where the spin and phonon spectra cross. 
Note that near the $\Gamma$-point longitudinal and transverse phonons have different frequencies so a circularly-polarised phonon is not an eigenmode of the system and therefore would evolve as a complex superposition of these modes. 
A purely longitudinal excitation along [111] direction cannot generate $\mathbf{L}_\mathrm{ph}$ and fails to induce $M_z$ precession (see Suppl. Info. Fig.~S2). 
 In Fig.~\ref{fig:Gamma-P-xz} we present an example of switching 
 excited in $xz$-plane. Although the $xz$-plane excitation contains both longitudinal and transverse components, the frequency mismatch between the longitudinal and transverse modes prevents them from maintaining a stable relative phase over time. As a result, the mixed longitudinal–transverse contributions average out in time, while the two degenerate transverse modes preserve a coherent relative phase, yielding an angular momentum aligned with the wave-vector, i.e., $L_x\approx L_y\approx L_z$ (more details in Suppl. Info. S3.2).
As shown in Fig~\ref{fig:Gamma-P-xz}(a), 
the left-handed phonon (and not the right one as near the $P$-point ) is absolutely inefficient and is not able to pass angular momentum to magnetization precession and to induce a large magnetization motion. 
In Suppl. Info. we present more examples of magnetization switching. Particularly, we investigate a pure transverse excitation in the plane perpendicular to the k-vector directed along [111] direction in Figs.~S3 and S4. 

\begin{figure}[!t]
\includegraphics[width=\linewidth]{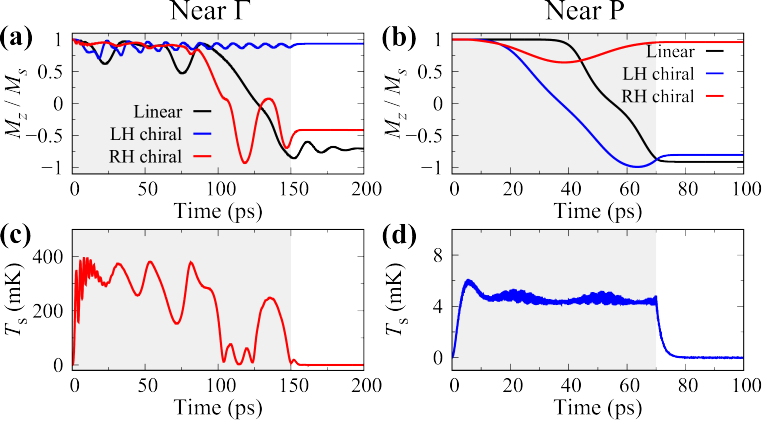}
\caption{\label{fig:Mz} Chirality-dependent magnetization switching dynamics (a, b) and spin temperature (c, d) driven by THz phonon excitation ($xz$-plane excitation). (a, c)\,Excitation near $\Gamma$-point with excitation force strength of $f_\text{0} = 0.05$ for both linear and chiral phonons. (b, d)\,Excitation near $P$-point with $f_\text{0} = 0.06$ for both linear and chiral phonons. The spin temperatures plotted in panels (c) and (d) correspond to the RH and LH chiral phonon excitations, respectively.}
\end{figure}

Comparatively to linear phonon excitation, a much larger phonon angular momentum (with opposite signs for the two chiralities) is directly injected into the system with circularly-polarised phonons, see Suppl. Info. Fig.~S6. For chiral phonons, this large constant phonon angular momentum arises in the system in the timescale below 10\,ps ( as compared to the linear excitation where its development needs at least 50\,ps).   Secondly. we note that the linear phonon is a sum of the two phonons with opposite chiralities, each with half of the  amplitude. This explains that the system switches both with a linear and a chiral phonon since the latter is contained in the linear excitation and one of them is not only inefficient but acts in the opposite direction. Both described effects lead to a faster angular momentum transfer and faster switching with chiral phonons, illustrated in Fig.~\ref{fig:Mz} and Fig.~S6.  
Most remarkable is  that  orders of magnitude  larger spin temperature accompanies the switching event with excitation near $\Gamma$ point in comparison with that near $P$-point, as shown in Figs~\ref{fig:Mz}(c) and \ref{fig:Mz}(d). Indeed, near the spin-phonon crossing point the elastic energy is not only passed to the magnetization switching itself but also induces a strong excitation of spinwaves (forbidden near the $P$-point).

To conclude, our results demonstrate that since spin dynamics have a well-defined rotational sense, only one phonon chirality effectively couples to the magnetization precession. Switching by chiral phonons is possible at different points of the spectrum. The switching efficiency is explained by a very large effective Barnett field reaching hundreds of Tesla at THz frequencies and destabilizing spin precession with one of the phonon polarisations. 
While switching with both linear and angular phonons is feasible with ultrafast THz pulses, the transfer of angular momentum with chiral phonons is very efficient and thus the switching is faster.   While it is believed that the most efficient switching should occur near the crossing  point of the two spectra, this excitation produces also a large energy transfer to the spin system, losing energy in spinwave excitations. Our results point  to the possibility of pure magneto-elastic mechanism of chiral phonon-induced  magnetization switching  without the need to evoke the quantum-mechanical effective fields coming from circular charge motion. As was  suggested theoretically~\cite{Matsuo2013Mechanical}, the mechanism discussed here can also act as a source of spin currents via the so-called spin-rotation coupling and additionally assist magnetization switching in magnetic multilayers.  

\textit{Acknowledgments}
M.Y. and W.H. acknowledge support from National Natural Science Foundation of China (Nos. 12302134, 12272173, and 11902150) and NUAA Ph.D. Student Short-Term Overseas Study Funding (241201DF01). O.C.-F. acknowledges grant PID2022-137567NB-C21 and  the Centres of Excellence program Severo Ochoa (Grant CEX2024-001445-S), both funded by MCIN/AEI/10.13039/501100011033. M.S., W.H. and O.C.-F. acknowledge the European COST Action CA23136 CHIROMAG. M.H.A and M.S. gratefully acknowledge funding from SOE Student Experience Internship from the University of Manchester and support from prof Thomas Thomson and the NEST group at the University of Manchester. This work was partially supported by the High Performance Computing Platform of NUAA.

\bibliography{reference}

\end{document}